\documentclass[twocolumn,showpacs,preprintnumbers,amsmath,amssymb]{revtex4}
\usepackage{longtable}
\usepackage{graphicx}% Include figure files
\usepackage{dcolumn}% Align table columns on decimal point
\usepackage{bm}% bold math
\usepackage{color}
\begin{document}

\preprint{APS/123-QED}

\title{Experimental investigation of a control scheme for a zero-detuning resonant sideband extraction interferometer
for next-generation gravitational-wave detectors }% Force line breaks with \\

\author{Fumiko Kawazoe}
 \email{fumiko.kawazoe@aei.mpg.de}
 %\altaffiliation[Also at ]{Physics Department, Ochanomizu University.}%Lines break automatically or can be forced with \\
\author{Akio Sugamoto}%

\affiliation{Ochanomizu University, 2-1-1 Otsuka, Bunkyo-ku,
Tokyo 112-8610, Japan\\
}%

\author{Volker Leonhardt, Shuichi Sato, Toshitaka Yamazaki, Mitsuhiro Fukushima, Seiji Kawamura}
\affiliation{ National Astronomical Observatory of Japan, 2-21-1
Osawa, Mitaka-shi, Tokyo
181-8588, Japan% with \\
}%

\author{Osamu Miyakawa}
\affiliation{LIGO Laboratory, California Institute of Technology,
Pasadena, CA
91125, USA % with \\
}%

\author{Kentaro Somiya}
\affiliation{ Max-Planck-Institut f$\ddot{u}$r Gravitationsphysik,
Am M$\ddot{u}$hlenberg 1, 14476 Potsdam, Germany
}%
\author{Tomoko Morioka}
\affiliation{ University of Tokyo, Kashiwa, Chiba 277-8582, Japan % with \\
}%
\author{Atsushi Nishizawa}
\affiliation{ Graduate School of Human and Environmental Studies, Kyoto University, Kyoto 606-8501, Japan% with \\
}%
\date{\today}% It is always \today, today,

\begin{abstract}
Some next-generation gravitational-wave detectors, such as the American Advanced LIGO project and the Japanese LCGT project, plan to use power recycled resonant sideband extraction (RSE) interferometers for their interferometer's optical configuration. A power recycled zero-detuning (PRZD) RSE interferometer, which is the default design for LCGT, has five main length
degrees of freedom that need to be controlled in order to operate a gravitational-wave detector. This task is expected to be very challenging because of the complexity of optical configuration. A new control scheme for a PRZD RSE interferometer has been developed and tested with a prototype interferometer. The PRZD RSE interferometer was successfully locked with the control scheme. It is the first experimental demonstration of a PRZD RSE interferometer with suspended test masses. The result serves as an important step for the operation of LCGT.
%Valid
%PACS numbers may be entered using the \verb+\pacs{#1}+ command.
\end{abstract}

\pacs{04.80.Nn, 42.60.Da, 95.55.Ym}% PACS, the Physics and Astronomy
\maketitle

\section{\label{sec:level1}Introduction\protect\\ }

%This sample document demonstrates proper use of REV\TeX~4 (and
%\LaTeXe) in mansucripts prepared for submission to APS journals.
%Further information can be found in the REV\TeX~4 documentation
%included in the distribution or available at
%\url{http://publish.aps.org/revtex4/}.
%
%{\color{blue}{We can change colors for emphasis}},
%{\color{green}{but}} {\color{cyan}{who is going pay for the ink?}}
%
%When commands are referred to in this example file, they are always
%shown with their required arguments, using normal \TeX{} format. In
%this format, \verb+#1+, \verb+#2+, etc. stand for required
%author-supplied arguments to commands. For example, in
%\verb+\section{#1}+ the \verb+#1+ stands for the title text of the
%author's section heading, and in \verb+\title{#1}+ the \verb+#1+
%stands for the title text of the paper.
%
%Line breaks in section headings at all levels can be introduced using
%\textbackslash\textbackslash. A blank input line tells \TeX\ that the
%paragraph has ended. Note that top-level section headings are
%automatically uppercased. If a specific letter or word should appear in
%lowercase instead, you must escape it using \verb+\lowercase{#1}+ as
%in the word ``via'' above.
Presently several laser interferometer gravitational-wave detectors are in operation in the United States, (LIGO\cite{sigg06}), in Europe, (GEO600\cite{lueck06}, and VIRGO\cite{virgo06}), and in Japan, (TAMA300\cite{ando05}).  In addition to the present detectors, there are plans to upgrade them to next-generation interferometers. Amongst them are Advanced LIGO\cite{frit03} and LCGT\cite{kuroda99}, which plan to use the power recycled resonant sideband extraction (RSE) technique to enhance detector sensitivities.

Despite the great advantage of being able to achieve better sensitivity by avoiding problems of thermal absorption by substrates, the RSE configuration poses a more difficult challenge in controlling the interferometer in order to use it as a gravitational-wave detector due to the increased number of degrees of freedom (DOF) that need to be controlled. Therefore, designing and demonstrating a control scheme as simple as possible are vital before the technique is adapted in large-scale interferometers such as LCGT. Recently a control of a power recycled detuned RSE has been demonstrated with a prototype experiment on the 40m interferometer at Caltech \cite{miyakawa06}, and a zero detuned signal recycling has been demonstrated on the GEO600 detector \cite{hild07}. The difference between the RSE and the signal recycling is explained in detail in a paper such as \cite{sato07}. 

We have developed a novel control scheme for LCGT \cite{sato07}, and have carried out experimental work\cite{kawazoe06}. This experiment aims to control a PRZD RSE interferometer using the scheme, and to measure the sensing matrix and compare it with modeling. The control scheme is described in section II, the experimental results are
presented in section III, the results are shown in section IV, discussions are presented in section V, and finally the conclusion is presented in section VI.

\section{\label{sec:level1}Control scheme\protect\\ }
The control scheme consists mainly of two parts; the signal extraction scheme and the lock sequence.
\subsection{\label{sec:level1}Signal extraction scheme\protect\\ }
The power recycled RSE interferometer has five DOFs to be controlled as shown in Fig. \ref{Degrees of freedom in the power recycled RSE interferometer}. They are the average length and the differential length of the two Fabry-Perot (FP) arm cavities, $L_+$, and $L_-$, respectively, as indicated by arrows with solid lines, the average length and the differential length of the power recycling cavity  (PRC), $l_+$, and $l_-$, respectively, as indicated by arrows with short dashed lines, and the average length of the signal extraction cavity (SEC), $l_{\text{s}}$, as indicated by arrows with long dashed lines.

\begin{figure}[htbp]
\includegraphics[width=21pc]{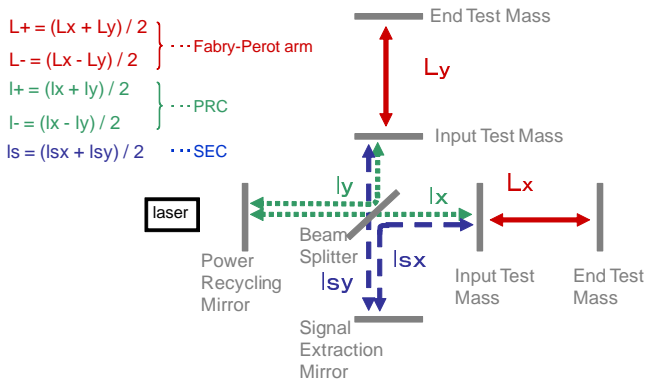}
\caption{\label{Degrees of freedom in the power recycled RSE
interferometer}Degrees of freedom in the power recycled RSE
interferometer.}
%\end{minipage}\hspace{2pc}%
\end{figure}

It is known from experience with present detectors which use the
Pound-Drever-Hall (PDH) method \cite{PDH} that the arm cavities are relatively
easy to control with clean control signals as the arm cavities have
high finesse. On the other hand, it is expected to be quite
challenging to obtain clean control signals of the central part of
the RSE (i.e. $l_+$, $l_-$, and $l_{\text{s}}$), because the resonant conditions of the light fields inside
the central part will be strongly affected by both the PRC and the
SEC. Thus the length-sensing scheme has to be designed in such a way
that it manipulates the resonant conditions of the light inside by
properly designed cavity lengths and sideband frequencies.

The outline of the length-sensing scheme is as follows: It is based
on the PDH method. The FP arm cavity lengths are controlled with a
single modulation-demodulation technique and the central part of the
RSE is controlled with a double modulation-demodulation technique
with amplitude modulation (AM) sidebands and phase modulation (PM)
sidebands. By using the double modulation-demodulation technique,
the control signals for the central part will be affected very
little by the signals derived from the carrier which are dominated
by the FP arm cavities. This is because both the AM and the PM
modulation sideband frequencies are designed in such a way that they
are not resonant in the FP arm cavities; thus it decouples the FP
arm cavities and the central part.

\subsubsection{The central part}
The central part is designed so that the AM and the PM sidebands behave in the following way: The lengths of two paths that compose the Michelson interferometer have a macroscopic asymmetry such that when the carrier interferes destructively at the dark port (DP), the AM sidebands interfere constructively at the bright port (BP) and destructively at the DP, while the PM sidebands interfere destructively at the BP and constructively at the DP. Thus the AM sidebands ``reflect completely" from the Michelson part while the PM sidebands ``transmit completely" through the Michelson part. This condition is met when the round trip Michelson asymmetry length is designed to be equal to $(2m+1)c/2f$ $(m=0,1,2,...)$ for the PM sidebands, where $c$ is the speed of light, and $f$ is the modulation freqency,  and integer multiple of $c/f$ for the AM sidebands. In our design it is $3c/f$ for the AM sidebands and $c/2f$ for the PM sidebands. The two cavities' macroscopic lengths are designed so that the AM sidebands resonate inside the PRC and the PM sidebands resonate inside the compound cavity made of the PRC and the SEC. This enables the PM sidebands to be sensitive to the length of the SEC while the AM sidebands are not affected by the SEC length, thus ensuring independent control signals for $l_+$ and $l_{\text{s}}$.

\subsubsection{The whole RSE}
\begin{figure}[htbp]
\includegraphics[width=19pc]{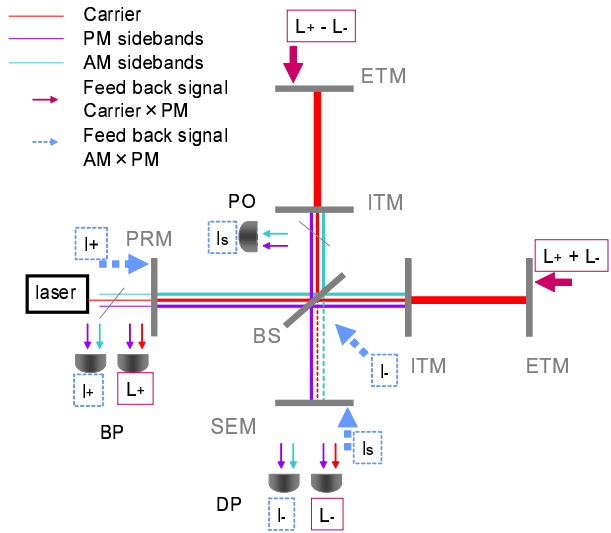}
\caption{\label{Control of the whole interferometer}Control of the
whole interferometer.}
\end{figure}

Figure \ref{Control of the whole interferometer} shows how the control signals of the prototype RSE interferometer are obtained. The $L_+$ control signal is obtained at BP and is fedback to the end test masses (ETM), the $L_-$ control signal is obtained at DP and is fedback to the ETMs. The $l_-$ control signal is obtained at DP and is fedback to the beam splitter (BS), the $l_+$ control signal is obtained at the BP and is fedback to the power recycling mirror (PRM), and the $l_{\text{s}}$ control signal is obtained at the pick-off port (PO) and is fedback to the signal extraction mirror (SEM).

Table \ref{Theoretical matrix} shows the theoretical length sensing signal matrix. The signals are at DC. It is calculated with parameters used for the prototype RSE interferometer, of which there will be a detailed explanation in section III. Demodulation phases are chosen so that each main signal (i.e. the signal that should be obtained at the corresponding detection port) is maximized. The first from the left column shows the detection ports; SD and DD stand for single/double modulation-demodulation, respectively, and the top first row shows the DOFs. The values along the same row are normalized by each main signal. The $L_+$ and $L_-$ signals are dominant at BP and DP, respectively, due to the designed high finesse of the Fabry-Perot arms, indicating they are relatively easy to obtain. On the other hand, it is obvious from the matrix that $l_+$ and $l_\text{s}$ signals mix at both BP and PO. However, the degree to which they mix is such that linearly independent signals can be obtained. Also $L_{+/-}$ signals that mix with the main signals on the DD systems are relatively small due to the fact that the carrier is not used for them. \footnote{One thing to note is that the amplitude of the $L_+$ signal which mixes at the PO(DD) system is comparable to the main signal $l_\text{s}$. Simulation demonstrates that this is because the macroscopic length of the arm cavity is such that the AM sidebands are not very far from the resonance peaks of the cavity. When choosing the length the size of the existing camber limited the freedom of choosing the best possible design, and so the optimum design is a compromising one. Therefore the size of the unwanted signal can be
reduced by choosing the right length of the FP arm cavity, thus it will not be a problem in a real detector such as LCGT.} This is necessary to bring the interferometer from an uncontrolled state to a controlled state. Therefore the designed control scheme is a promising approach for acquiring the control.

 \begin{table}[htbp]%[H] add [H] placement to break table across pages
 \caption{\label{Theoretical matrix}Theoretical matrix}
 \begin{ruledtabular}
 \begin{tabular}{ccccccc}
&$L_+$&$L_-$&$l_+$&$l_-$&$l_\text{s}$\\
\hline
BP(SD)&1&8.2$\times10^{-6}$&-2.6$\times10^{-2}$&6.4$\times10^{-4}$&1.3$\times10^{-2}$\\
DP(SD)&-9.2$\times10^{-9}$&1&5.9$\times10^{-9}$&1.3$\times10^{-2}$&8.6$\times10^{-9}$\\
BP(DD)&-4.9$\times10^{-2}$&-1.1$\times10^{-4}$&1&-8.6$\times10^{-3}$&-5.3$\times10^{-1}$\\
DP(DD)&-1.0$\times10^{-4}$&7.6$\times10^{-2}$&1.4$\times10^{-3}$&1&1.1$\times10^{-5}$\\
PO(DD)&-1.5$\times10^{-1}$&-1.2$\times10^{-2}$&1.1&-2.2$\times10^{-2}$&1\\
 \end{tabular}
 \end{ruledtabular}
 \end{table}
%\end{table*}

\subsection{\label{sec:level1}Lock sequence\protect\\ }
Up until this point we had assumed that the interferometer is close to its operating point; therefore feedback signals for all the DOFs are always present for the control system to work properly. In reality the uncontrolled interferometer is not near its operating point most of the time, therefore feedback signals for all the DOFs
are almost never present simultaneously. In order to successfully lock the interferometer, a sequence of locking needs to be established. Here we show one possible order of locking the whole interferometer with simulation work results. FINESSE \cite{FINESSE} is used for the calculation. The sequence is as follows:
\begin{enumerate}
\item Lock the central part
\begin{enumerate}
\item Lock $l_-$ DOF
\item Lock $l_+$ DOF
\item Lock $l_\text{s}$ DOF
\end{enumerate}
\item Lock the FP arm cavities
\end{enumerate}
Figure \ref{$l_{-}$ error signal with SEM and PRM free} shows the feedback signal of the $l_{-}$ DOF when the other two DOFs in the central part are free of control. Here the FP arms are not assumed to be present for simplicity. The plots are contour plots of the error signal for the $l_{-}$ DOF when the other two DOFs are uncontrolled, i.e. the microscopic position of the SEM expressed in the phase gained by the light (hereinafter called phi) is scanned
from its operating point by 0, 30, 60, and 90 degrees as indicated by the vertical arrow, while the PRM's phi is scanned from its operating point by -90 to 90 degrees in the direction of the y-axis. The x-axis shows the phi of the $l_-$ DOF, with the $\text{phi}=0$ being the operating point. On the right of each plot there is a color bar that shows the signal amplitude.

The $l_-$ DOF can be locked to where the amplitude of the error signal changes its sign at the so-called zero crossing. A clear vertical line of a zero crossing is present at the operating point
in the top figure when the $l_+$ and the $l_\text{s}$ DOFs happen to be at their operating points, (i.e. their phi are zero). The polarity of the zero crossing needs to be taken into account; in this case the right polarity is the one that goes from plus to minus as the phi is moved from minus to plus. With the wrong polarity the error signal does not guide the DOF to be locked at the operating point. A zero crossing line at the operating point having the wrong polarity is seen as the phi of the $l_\text{s}$ DOF is scanned between 0 to 90 degrees. One such example is marked by the ellipse in the figure. According to the simulation results the chance of such a line being present is $\sim$ 30\% at the maximum. This indicates that most of the time (with a minimum chance of being $\sim$ 70\%) when the other two DOFs are completely free of control, (i.e. their detunings are scanned independently from each other), the $l_{-}$ DOF can be locked at its operating point.

Then, with the assumption that the $l_-$ is now locked at its operating point, the feedback signal of the $l_{+}$ DOF is shown in Fig. \ref{$l_{+}$ error signal with SEM free and $l_{-}$ in lock} when the $l_{s}$ is still free of control so it could be anywhere in the y-axis. In the same manner as in Fig. \ref{$l_{-}$ error signal with SEM and PRM free}, it is clear that a zero crossing for the $l_{+}$ is present very close to its operating point regardless of the $l_\text{s}$'s phi detuning. Therefore the $l_+$ DOF can be locked at its operating point. When this is the case, the zero crossing line of the $l_-$ DOF is always present as shown in the top plot of Fig. \ref{$l_{-}$ error signal with SEM and PRM free}, so the lock of the $l_-$ DOF becomes stable.

Next, along with the assumption that the $l_-$ is locked, there is a new assumption that the $l_+$ is locked at its operating point, a feedback signal of the $l_{s}$ DOF is shown in Fig. \ref{$l_{s}$ error signal with $l_{+}$ and $l_{-}$ in lock}. The figure shows that a zero crossing point is present at its operation point as marked by the arrow. In this way the central part is locked at its operating point.

This is a good approximation because the chance that the carrier is resonant inside the FP arm cavities and thus disturbs the central part length sensing signals is very small because of the high finesse, and the chance that the sidebands are accidentally resonant inside the FP arm cavities is very small for the same reason.

Next the FP arms are locked by acquiring the lock of the individual FP arm whose control signal is always present and dominant at each detection port for each FP arm cavity. Once the two FP arm cavities are locked, the control servos can be switched to those of the $L_{-}$ and $L_{+}$ DOF to change them into the controls of $L_{-}$ and $L_{+}$ DOFs simultaneously. Therefore the FP arms are locked. Thus the whole RSE interferometer can be locked in this order.

\begin{figure}[htbp]
\includegraphics[width=20pc]{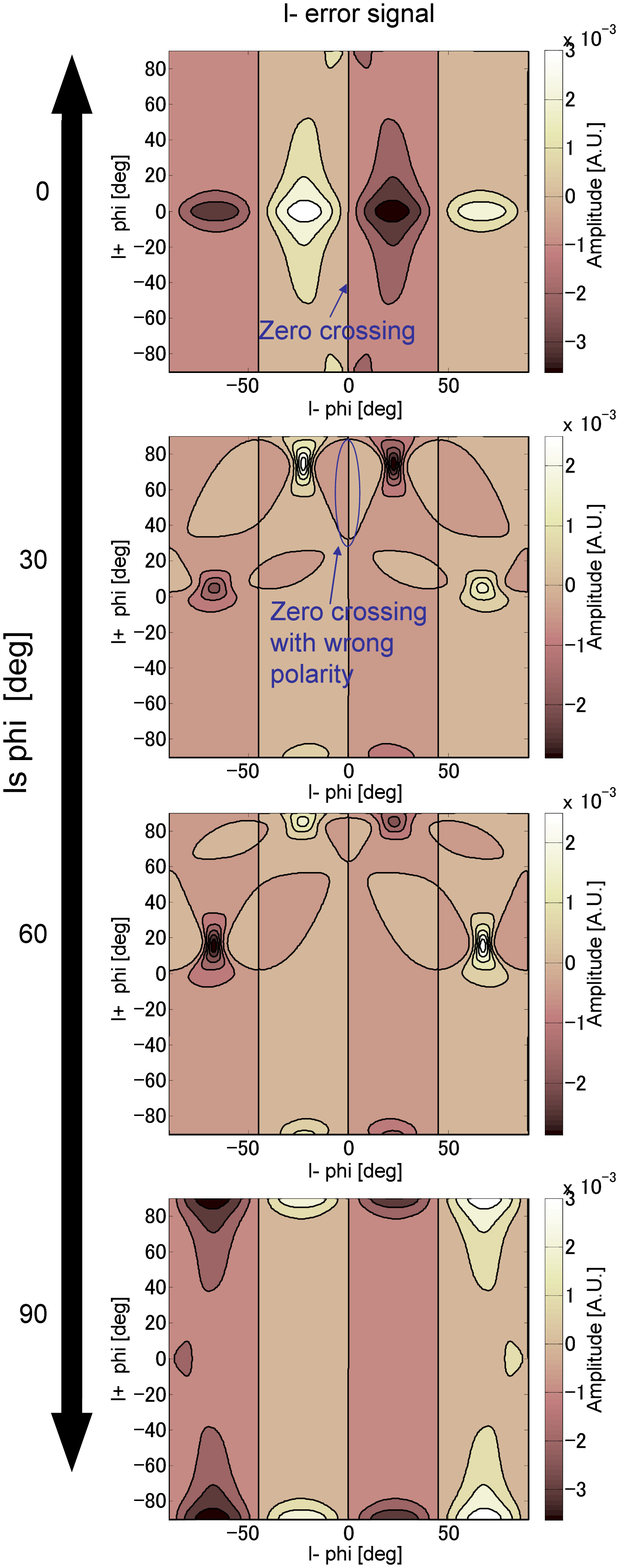}
\caption{\label{$l_{-}$ error signal with SEM and PRM free}$l_{-}$
error signal with SEM and PRM free.}
\end{figure}
\begin{figure}[htbp]
\includegraphics[width=20pc]{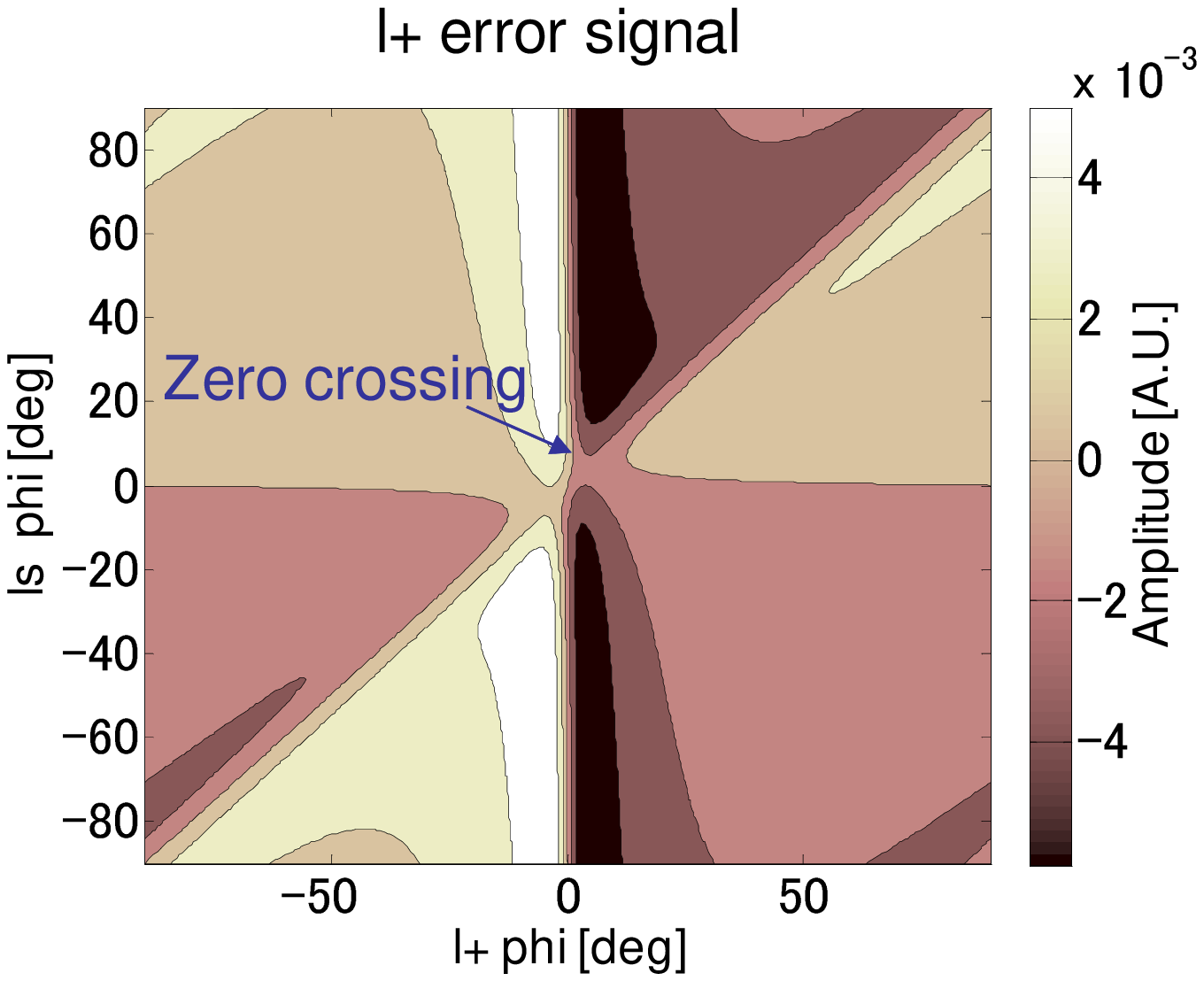}
\caption{\label{$l_{+}$ error signal with SEM free and $l_{-}$ in
lock}$l_{+}$ error signal with SEM free and $l_{-}$ in lock.}
\end{figure}
\begin{figure}[htbp]
\includegraphics[width=20pc]{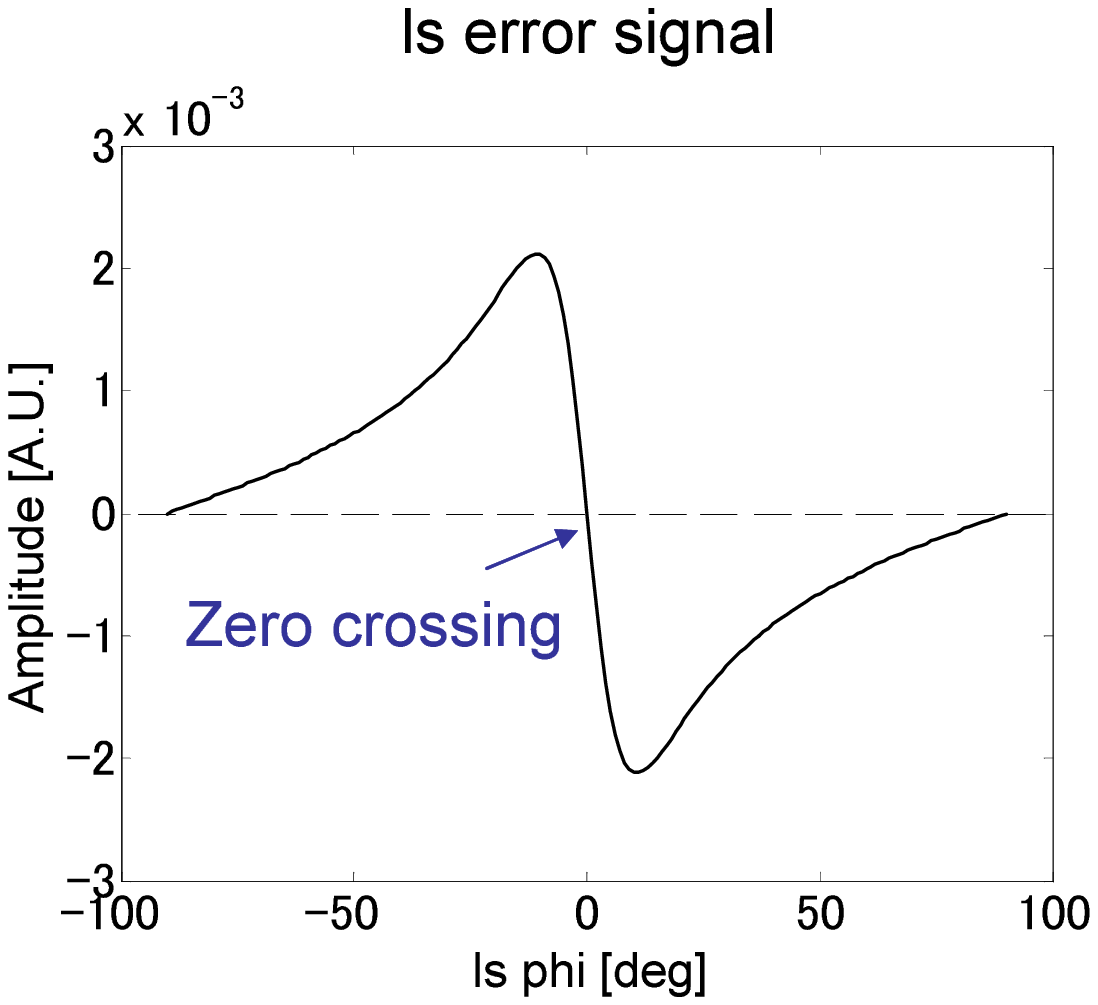}
\caption{\label{$l_{s}$ error signal with $l_{+}$ and $l_{-}$ in
lock}$l_{s}$ error signal with $l_{+}$ and $l_{-}$ in lock.}
\end{figure}
\section{\label{sec:level1}Experimental demonstration of a control of the PRZD RSE\protect\\ }
\subsection{\label{sec:level1}4m Prototype interferometer\protect\\ }
Figure \ref{Optical layout of the prototype interferometer} shows the optical layout of the prototype interferometer. A light at 1.064 $\mu$m from a Nd:YAG laser enters a Mach-Zehnder interferometer (MZ) where two Pockels cells are placed in different paths of the MZ and it is phase modulated at 17.25 MHz and amplitude modulated at 103.25 MHz. Light power at the output of the MZ is 90 mW with modulation depths of approximately 0.15 rad for both frequencies. The FP arm cavity length is 4.15m and is formed by a flat input mirror and a curved end mirror with a radius of curvature of 6m. Two mode-matching lenses are used for the FP arm cavities and an additional two are used to compensate for the mismatch due to the relatively large Michelson macroscopic length asymmetry (4.35 m). A coupling ratio between the incident beam and the fundamental mode of the FP arm cavity of more than 98\% is achieved. The finesse of the cavity is approximately 120. All of the seven test masses (diameter 2.54cm) are suspended as double
pendulums (height 30cm) to suppress the mirror motion at frequencies above the resonant frequency (at about 1.5Hz). The test mass motion around its resonant frequency is damped by an eddy-current damping
system. The length control is performed via coil-magnets actuators.
\begin{figure}[htbp]
\includegraphics[width=20pc]{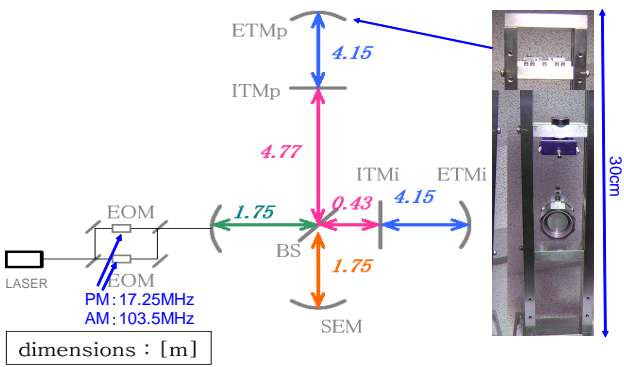}
\caption{\label{Optical layout of the prototype
interferometer}Optical layout of the prototype interferometer.}
\end{figure}

\subsection{\label{sec:level1}Lock Sequence\protect\\ }
The lock of the RSE has been successfully demonstrated. The sequence
that has been determined with the simulation work is used; the
central part first then the FP arm cavities next. The
direct lock of the FP arm cavities' two DOFs, the common, $L_+$, and
the differential, $L_-$, has been realized without one additional
step (locking the individual arms with each control signal, then
switch the control servos for those of the common and the
differential control). This is because often one of the two FP arm
cavities is locked with the control signal for the $L_+$ DOF first,
and as soon as the other FP arm cavity's DOF is close to its
operating point the $L_-$ DOF is controlled, consequently switching
the $L_+$ control signal to lock the $L_+$ DOF. Typically the time
interval between the lock of one FP arm cavity and both cavities is
less than a second.

\begin{figure}[htbp]
\includegraphics[width=20pc]{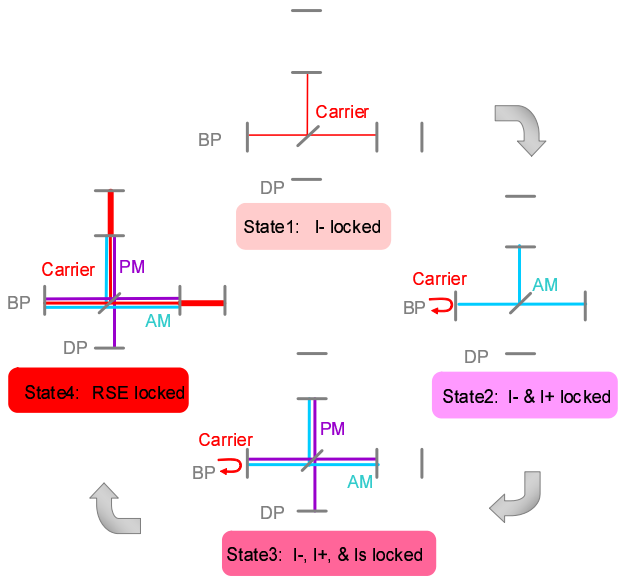}
\caption{\label{Lock states of the interferometer}Lock states of the
interferometer.}
\end{figure}

We define several lock states, which are shown in Fig. \ref{Lock states of the interferometer}. Each state is indicated as follows:
\begin{itemize}
\item State 0 : The interferometer is uncontrolled.
\item State 1 : Michelson is locked to dark fringe at the DP.
\item State 2 : Power-recycled Michelson is locked.
\item State 3 : Central part is locked.
\item State 4 : Tuned RSE is locked
\end{itemize}

\begin{figure}[htbp]
\includegraphics[width=23pc]{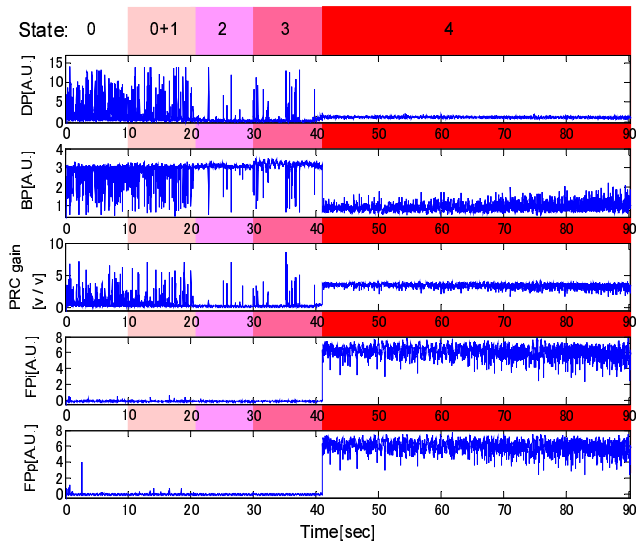}
\caption{\label{DC power at various ports}DC power at various
ports.}
\end{figure}

In state 2, the carrier is anti-resonant inside the PRC. In order for the carrier to be resonant inside the PRC, the phase needs to be shifted by $\pi$ inside the PRC, which happens when the carrier is resonant inside the FP arm cavities as in state 4. Figure \ref{DC power at various ports} shows the DC power detected at various ports, (i.e. DP, BP, PO, transmitted port for the inline FP arm cavity and for the perpendicular FP arm cavity). Each lock state is separated by boxes with colors specified in Fig. \ref{Lock states of the interferometer}, and on top of each box the state number is shown. The start time of each state is defined as the time when the switch of a servo loop of each DOF is turned on. Typically each DOF is locked within a fraction of a second after the servo loop is switched on. Between the time 10 and 20 sec, the $l_-$ DOF is not locked the whole time. This is because the carrier is not
yet anti-resonant inside the PRC and disturbs the control signal of the $l_-$ DOF. When the $l_+$ servo loop is switched on at time 20 sec, the carrier is anti-resonant inside the PRC and the $l_-$ DOF is completely locked as well as the $l_+$ DOF. The difference between states 2 and 3 is not obvious from the DC signals, due to the relatively small size of the modulation sidebands compared to that of the carrier. The sharp peaks seen between time 20 and 40 sec can be explained as follows: The carrier light is sometimes resonant in one of the two arms, consequently adding a relative phase shift of $\pi$ between the two beams interfering at the BS. This will switch the DP into BP, and vice versa, so the DP is no longer dark. Note that the corresponding peaks in  light transmitted from the FP arm cavities are too small to be seen in the plots.

In order to verify the locking status, resonant conditions for the
two sets of sidebands are monitored with optical spectrum analyzers
(OSA) placed at the DP and the PO. Figure \ref{Sideband resonant
peaks inside the PRC and the SEC} shows the output of the optical
spectrum analyzers. The upper three plots show the output power at
the PO and the bottom three show the output power at the DP. Colored
boxes indicate the lock state, as specified in Fig. \ref{Lock states
of the interferometer}. In state 1, neither the AM nor the PM
sidebands are resonant. In state 2, the AM sidebands are resonant
inside the PRC, thus there are resonant peaks of the AM sidebands
detected at the PO. The PM sidebands are not yet resonant. In state
3, the PM sidebands are resonant inside the compound cavity made of
the PRC and the SEC thus there are resonant peaks of the PM
sidebands detected at the DP. Each cavity length is controlled
throughout the locking process.

\begin{figure}[htbp]
\includegraphics[width=22pc]{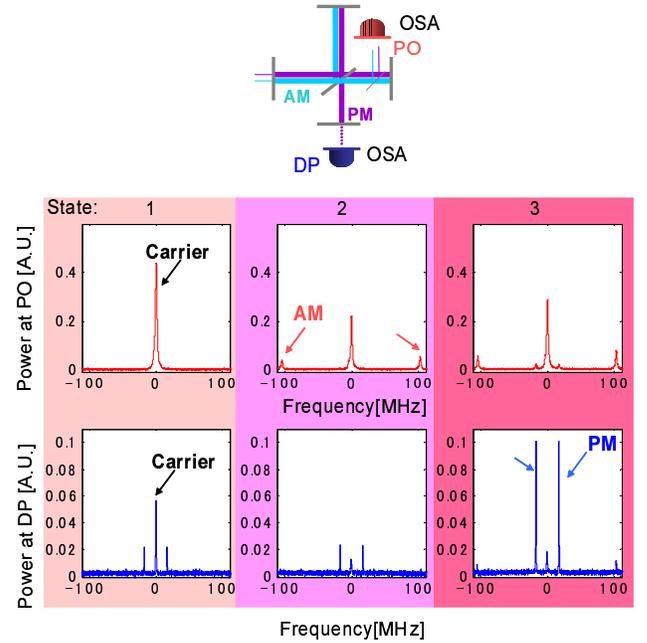}
\caption{\label{Sideband resonant peaks inside the PRC and the
SEC}Sideband resonant peaks inside the PRC and the SEC.}
\end{figure}

Table \ref{Measured normalized matrix} shows the measured length sensing signal matrix, normalized by the main signals at each detection port. Demodulation phases are tuned to maximize the main signal at each detection port. A sinusoidal signal is applied to a feedback path of each DOF to move the test masses. The signal frequency (at 2.2 kHz) is well above the unity gain frequency of all the control servo so that loop gains do not have to be taken into account. The frequency is also well below the cut-off frequency of the optical response of the RSE (at about 1 MHz), so the values can be directly compared with the theoretical DC values.

The general pattern of the measured matrix is in good agreement with the theoretical one; the $L_{+/-}$ signals dominate at their detection ports, the $l_+$ and the $l_\text{s}$ mix at BP and PO but the signals are linearly independent, and the $L_{+/-}$ signals that mix at DD systems are relatively small. In general, unwanted signals other than the one that should be obtained at the corresponding detection ports tend to be greater than the theoretical values due to present imperfections such as the unwanted carrier light at the DP. It is not essential for this experiment where agreement in the overall feature of the matrix is of importance in order to verify the control scheme.
\begin{table}[htbp]
\caption{\label{Measured normalized matrix}Measured normalized
matrix}
\begin{ruledtabular}
\begin{tabular}{cccccccc}
&$L_+$&$L_-$&$l_+$&$l_-$&$l_\text{s}$\\
\hline
BP(SD)&1&4.5$\times10^{-2}$&4.8$\times10^{-2}$&-1.9$\times10^{-2}$&-1.3$\times10^{-2}$\\
DP(SD)&-1.9$\times10^{-1}$&1&-2.1$\times10^{-3}$&-4.9$\times10^{-2}$&-1.5$\times10^{-3}$\\
BP(DD)&-1.2$\times10^{-1}$&6.6$\times10^{-3}$&1&1.8$\times10^{-2}$&-5.0$\times10^{-2}$\\
DP(DD)&-1.1$\times10^{-1}$&-3.1$\times10^{-1}$&1.1$\times10^{-1}$&1&1.2$\times10^{-1}$\\
PO(DD)&-1.1&2.2$\times10^{-1}$&4.0$\times10^{-1}$&5.1$\times10^{-1}$&1\\
\end{tabular}
\end{ruledtabular}
\end{table}

\section{\label{sec:level1}Results\protect\\ }
The lock of the PRZD RSE has been successfully demonstrated with the prototype interferometer. It is the first experimental demonstration of a PRZD RSE interferometer with suspended test masses. In parallel the control sequence has been established with a simulation. The measured signal matrix showed good agreement with modeling.

\section{\label{sec:level1}Conclusion\protect\\ }
A new control scheme has been developed for LCGT, and a PRZD RSE
interferometer has been successfully controlled with the scheme.
This result has shown that the LCGT can use a PRZD RSE
interferometer as its optical configuration and use our control
scheme to lock the interferometer. This is
one important step towards the successful operation of the
LCGT detector.
%Although the control scheme is shown to deteriorate the quantum
%noise limited sensitivity at around 20 to 100 Hz, a feed forward
%technique can be used to reduce the noise to meet the sensitivity
%requirement.
\begin{acknowledgments}
This research is supported in part by a Grant-in-Aid for Scientific
Research on Priority Areas (415) of the Ministry of Education,
Culture, Sports, Science, and Technology of Japan, and also
partially supported by the US National Science Foundation under
cooperative agreement PHY-0107417.
\end{acknowledgments}
\bibliography{RSE2}% Produces the bibliography via BibTeX.
%\bibliography{apssamp}

\end{document}